# Integrated RF/Optical Wireless Networks for Improving QoS in Indoor and Transportation Applications

Mostafa Zaman Chowdhury, Md. Tanvir Hossan, Moh. Khalid Hasan, and Yeong Min Jang
Dept. of Electronics Engineering, Kookmin University, Seoul, Korea
E-mail: mzaman@kookmin.ac.kr, mthossan@ieee.org, khalidrahman45@gmail.com, yjang@kookmin.ac.kr

*Abstract*—Communications based solely on radio frequency (RF) networks cannot provide adequate quality of service (QoS) for the rapidly growing demands of wireless connectivity. Since devices operating in the optical spectrum do not interfere with those using the RF spectrum, wireless networks based on the optical spectrum can be added to existing RF networks to fulfill this demand. Hence, optical wireless communication (OWC) technology can be an excellent complement to RF-based technology to provide improved service. Promising OWC systems include light fidelity (LiFi), visible light communication, optical camera communication (OCC), and free-space optical communication (FSOC). OWC and RF systems have differing limitations, and the integration of RF and optical wireless networks can overcome the limitations of both systems. This paper describes an LiFi/femtocell hybrid network system for indoor environments. Low signal-to-interference-plus-noise ratios and the shortage bandwidth problems of existing RF femtocell networks can be overcome with the proposed hybrid model. Moreover, we describe an integrated RF/optical wireless system that can be employed for users inside a vehicle, remote monitoring of ambulance patients, vehicle tracking, and vehicle-to-vehicle communications. We consider LiFi, OCC, and FSOC as the optical wireless technologies to be used for communication support in transportation, and assume macrocells, femtocells, and wireless fidelity to be the corresponding RF technologies. We describe handover management—including detailed call flow, interference management, link reliability improvement, and group handover provisioning—for integrated networks. Performance analyses demonstrate the significance of the proposed integrated RF/optical wireless systems.

## 1. Introduction

Currently, radio frequency (RF)-based technologies are widely used for all aspects of wireless connectivity. The RF band lies between 30 kHz and 300 GHz in the electromagnetic spectrum [1], and its use is strictly regulated [2]. Most RF communications employ frequencies below 10 GHz because of its favorable communication properties, and this portion of the spectrum is almost exhausted. It is anticipated that future demands for wireless data traffic cannot be met by RF wireless technologies alone [3]. Moreover, RF systems suffer from severe interference between nearby RF access points (APs). Recently, the use of the optical portion of the spectrum for wireless connectivity has attracted researchers as a promising approach for satisfying the rising demand for high-data-rate wireless services. Optical wireless communication (OWC) [1], [2], [4]-[9] using infrared (IR), visible light (VL), and ultraviolet (UV) spectra—is considered to be the best complement to the currently used RF-based communications, which are not sufficient to meet the growing demand. In principle, OWC can offload considerable data traffic from RF networks. A 100 Gbps data rate has already been demonstrated using OWC [3]. It can perform well for both indoor and outdoor communication services. A few significant advantages of OWC include (i) massive unregulated bandwidth, (ii) a very high level of security, (iii) very low power consumption, (iv) no licensing fee, and (v) no interference with the RF spectrum [3], [10]. Light fidelity (LiFi), visible light communication (VLC), optical camera communication (OCC), and free-space optical communication (FSOC) are considered as emerging OWC technologies. The important physical challenging issues for the OWC systems are: (i) sensitivity to blocking by obstacles, (ii) limited transmitted power, (iii) atmospheric loss, inter-cell interference, and (iii) interference from ambient light [11]. Since OWC technologies suffer from several limitations, the integration of RF and optical wireless systems seems likely to be an effective approach for achieving desirable services for upcoming fifth-generation (5G) communications and beyond. This requires coordination between the RF and optical wireless systems in order for them to work as a single system, and the integration can be accomplished with hybrid and/or coexisting networks. In a hybrid system, a wireless link is served by any existing alternative network, and both RF and optical wireless networks work together to provide wireless links in the coexisting system.

### 1.1. Integrated Systems

LiFi systems for indoor use provide a great opportunity to divert considerable data traffic from existing RF systems. However, these systems suffer from several limitations: (i) LiFi-to-LiFi handover problems, due to the coverage hole between the serving LiFi and the target LiFi and (ii) very high levels of interference experienced by users in areas with overlapping LiFi coverage. These problems can be overcome by deploying RF femtocells in the same indoor area. However, the dense deployment of RF femtocells also suffers from several problems: (i) femtocell users receive huge interference due to signals received from neighboring many femtocells and (ii) femtocell edge users receive low level of signal power compared to high level of interference power from neighboring femtocells and different sources of noise. Hence, these users experience low signal-to-interference-plus-noise (SINR). These limitations can be significantly reduced by deploying LiFi along with the femtocells. Hence, a hybrid LiFi/femtocell system provides a good solution for indoor users.

The widely used RF-based wireless communications in transportation also suffer from several issues: a low level of quality of service (QoS) for vehicle users, high levels of interference, and low levels of link reliability. Nowadays, vehicle-to-vehicle (V2V) and vehicle-to-infrastructure (V2I) communications also employ OWC systems [9], [11]-[13]. However, communications using OWC technologies suffer from non-line-of-sight (NLOS) problems and environmental effects [2]. Hence, integrated RF/optical wireless technologies also provide the best approach for overcoming these limitations.

### 1.2. Literature Review

Only a few publications have considered hybrid systems for indoor wireless connectivity, and none has addressed a hybrid system for transportation. Most existing papers discuss the load-balancing issue. Also, most indoor solutions consider only wireless fidelity (WiFi) for RF networks, and RF femtocells have not been considered adequately. Consideration of femtocell networks is important, as femtocells use the same frequency spectrum as that of a macrocell network. Reference [14] reviews the existing research on the coexistence of WiFi and LiFi. The authors also discuss the research challenges regarding the coexistence issue. A fuzzy-logic-based handover technique for indoor LiFi and an RF hybrid network is presented in [15]. References [16] and [17] discuss the issue of AP selection in a hybrid VLC/RF network. Reference [18] evaluates the outage performance of the RF uplink in a hybrid VLC/RF system, considering the randomness of the positions of legitimate receivers and eavesdroppers, as well as light-energy harvesting. Reference [19] proposes a load-balancing method that focuses on optimizing network throughput over time. Reference [20] investigates vertical handover for heterogeneous VLC/RF systems. The authors formulate the problem as a Markov decision process and adopt a dynamic approach to obtain a tradeoff between the switching cost and delay requirement. In [21], the authors provide a delay analysis for the coexistence of omnidirectional small cells and directional VLC.

**1.3 Contribution of this Paper**

In the present study, we consider different OWC technologies, using RF femtocells, WiFi, and macrocell networks, to construct integrated systems. We discuss systems for indoor as well as outdoor applications. We describe integrated RF/optical wireless systems for two important cases: (i) wireless connectivity for indoor use and (ii) wireless connectivity for transportation. The main contribution of this paper can be summarized as follows:

- For the indoor case, we describe a LiFi/femtocell hybrid network, and discuss the limitations of each technology clearly.
- We describe a complete solution for mobility management and interference management by introducing zoning of the complete LiFi/femtocell coverage area.
- We also discuss a probability analysis for users in different zones, and propose an admission policy for newly originating and handover calls.
- We propose femto-access-point (FAP) idle-mode operation and the shifting of users between FAPs and LiFi APs.
- We propose a detailed handover protocol for call flow using the proposed hybrid LiFi/femtocell network system.
- We present integrated RF/optical wireless systems for the transportation case, and consider various applications.
- Interference management, the improvement of link reliability, and group handover provisioning are discussed in detail for the transportation case.

The rest of the paper is organized as follows. Section 2 presents an overview of OWC and femtocell systems and their respective channel models. The indoor hybrid LiFi/femtocell system layout, network selection, mobility management, and interference management are discussed in Section 3. Systems based on integrated RF/optical wireless technologies for transportation are presented in Section 4. Performance analyses are discussed in Section 5. Finally, Section 6 concludes the paper.

**2. System Overview**

**2.1 Overview of OWC and Femtocell Systems**

The most important OWC technologies are VLC, LiFi, OCC, and FSOC. Fig. 1 shows an illustrative deployment scenario for various OWC and RF technologies. All wired and wireless technologies are connected to each other through a core network. The various OWC and RF femtocell technologies are briefly discussed below.

Visible light communication (VLC) [22]-[24] uses light-emitting diodes (LEDs) or laser diodes (LDs) as transmitters and photodetectors (PDs) as receivers. This approach uses VL as the communication medium, which can provide communication, illumination, and localization. High-data-rate VLC systems have already been reported in several papers. VL is not harmful to humans, and offers substantial opportunities for applications in offices, homes, cars, trains, airplanes, and along the roadside.

LiFi [14], [25], [26] is a similar technology to WiFi, and provides high-speed wireless connectivity, along with illumination, using LEDs or LDs as transmitters and PDs as receivers. For the communication medium, it uses VL in the forward path and VL/IR/UV in the return path; it can provide communication, illumination, and localization. LiFi is a complete wireless networking system and supports seamless user mobility. It also supports point-to-multipoint and multipoint-to-point communications.

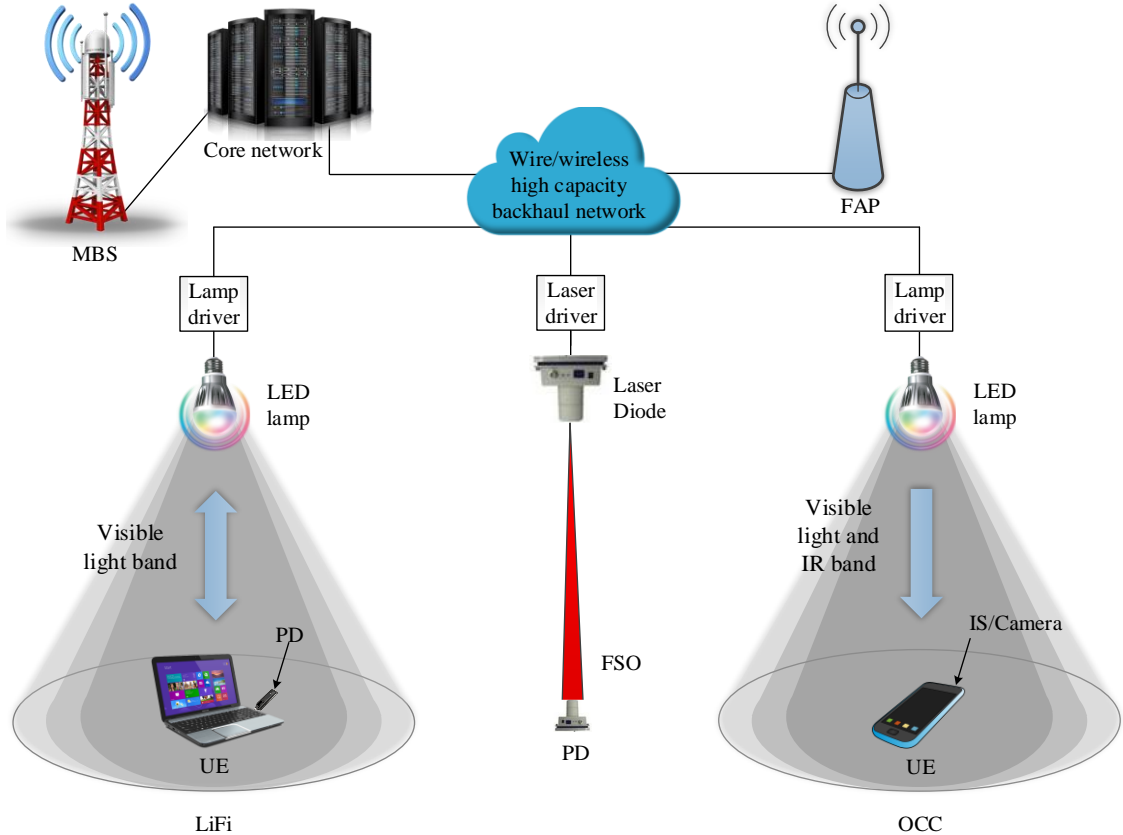

**Fig. 1**. Deployment scenario for various OWC and RF femtocell technologies.

OCC [12], [27], [28] uses LEDs and a camera [or image sensor (IS)] as the transmitter and receiver, respectively, and can use IR or VL as the communication medium. This technology can provide better performance, even in outdoor environments, so OCC systems can be used for indoor as well as outdoor applications. This technology is very promising for V2X (vehicle-to-vehicle, vehicle-to-infrastructure, and infrastructure-to-vehicle) connectivity.

FSOC [29]-[31] uses LDs as transmitters and PDs as receivers. This technology normally operates using IR as the communication medium. However, it can be also operated using VL and UV. It can provide very-long-distance (more than 10,000 km) communication with very high data rates. It has a wide range of applications, including backhaul connectivity for cellular networks, campus connectivity, disaster recovery, underwater communications, MAN-to-MAN/LAN-to-LAN connectivity, and ship-to-ship connectivity.

Femtocell [32]–[34] technology is widely deployed in subscribers' homes to provide high QoS. An FAP is a small-sized cellular base station deployed in a subscriber's home. FAPs are operated in the spectrum licensed for cellular-service providers. As femtocell networks coexist with macrocellular networks and share the same frequency band, the mitigation of interference between these two network types is an important issue. The effects of the resulting interference depend on the density of femtocells. Improper interference management can cause a significant reduction in system capacity and increase outage probability.

**2.2 Optical Channel Model**

The line-of-sight (LOS) optical channel gain $H_{i,j}$ between the $i^{\text{th}}$ optical AP and $j^{\text{th}}$ user is expressed as [15], [17]

$$H_{i,j} = \begin{cases} \dfrac{(m+1)A_P}{2\pi(l_{i,j}^2 + h^2)} g(\theta) T_s(\theta) \cos^m(\phi) \cos(\theta), & 0 \leq \theta \leq \Theta_F \\ 0, & \theta > \Theta_F \end{cases} \qquad (1)$$

where $m = \ln 2 / \ln(\cos\theta_{1/2})$ is the Lambertian index, $\theta_{1/2}$ is the half-intensity radiation angle, $A_P$ is the physical area of the PD, $g(\theta)$ is the optical concentrator gain, $T_s(\theta)$ denotes the optical-filter gain, $h$ is the height of the AP, $l_{i,j}$ is the horizontal distance between the $i^{th}$ optical AP and $j^{th}$ user, $\phi$ is the angle of irradiation, $\theta$ is the angle of incidence, and $\Theta_F$ is the half-angle of the field of view (FOV).

The optical concentrator gain can be expressed as

$$g(\theta) = \begin{cases} \dfrac{n^2}{\sin^2\Theta_F}, & 0 \leq \theta \leq \Theta_F \\ 0, & \theta > \Theta_F \end{cases} \qquad (2)$$

where $n$ is the refractive index.

The SINR for the $j^{th}$ user and the $i^{th}$ optical AP can be expressed as

$$SINR_{i,j} = \frac{(\Re P_{t,o} H_{i,j})^2}{N_0 B_o + \sum_{k \neq i}(\Re P_{t,o} H_{k,j})^2} \qquad (3)$$

where $N_0$ is the noise spectral density in units of area$^2$/Hz, $\Re$ is the optical to electrical conversion efficiency, $B_o$ is the system bandwidth of the optical AP, $P_{t,o}$ is the optical power transmitted by an optical AP, and $H_{k,j}$ is the channel gain between the $k^{th}$ interfering optical AP and $j^{th}$ user.

The achievable data rate between the $j^{th}$ user and $i^{th}$ optical AP can be expressed as

$$C_{i,j} = B_o \log_2(1 + SINR_{i,j}) \qquad (4)$$

## 2.3 RF Channel Model

The path loss for a macrocell is [35]

$$L_{macro} = 69.55 + 26.16 \log f_c - 13.82 \log h_b - a(h_m) + (44.9 - 6.55 \log h_b) \log d + L_{ow} \quad [\text{dB}] \qquad (5)$$

$$a(h_m) = 1.1(\log f_c - 0.7) h_m - (1.56 \log f_c - 0.8) \quad [\text{dB}] \qquad (6)$$

where $d$ is the distance, in km, between the macrocellular base station (MBS) and the user, $f_c$ is the central frequency in MHz, $h_m$ is the height of the user in m, $h_b$ is the height of MBS in m, and $L_{ow}$ represents the wall-penetration loss.

The propagation model for a femtocell is [36]

$$L_{femto} = 20 \log f + N \log z + 4q^2 - 28 \quad [\text{dB}] \qquad (7)$$

where $z$ is the distance, in m, between the FAP and the user, $f$ is the central frequency in MHz, $N$ is the distance power loss coefficient, and $q$ is the number of walls between the transmitter and receiver.

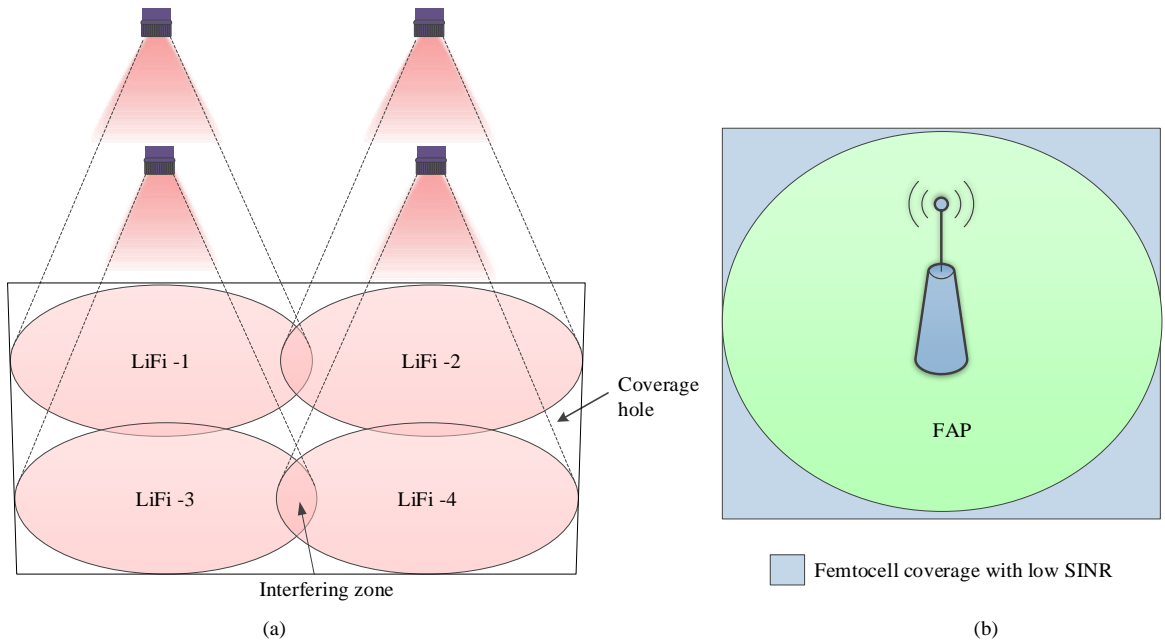

**Fig. 2.** Scenario illustrating the limitations of LiFi and RF femtocells. (a) Coverage hole and low SINR in an LiFi system. (b) Low-SINR coverage in a femtocell.

## 3. Hybrid LiFi/Femtocell System for Indoor Usage

### 3.1 Hybrid *LiFi/Femtocell Layout*

Both LiFi and femtocell technologies have limitations as well as advantages. The LiFi coverage area is comparatively very small; hence, the probability of handover failure due to the coverage hole is very high. Also, a user within an overlapped LiFi coverage area experiences a low SINR level. A dense deployment of RF femtocells causes higher levels of interference, especially for users at the edges. Fig. 2 shows a scenario for illustrating the limitations of LiFi and RF femtocell systems. Fig. 2(a) shows that when four LiFi APs are installed inside a room, there is a coverage hole due the circular nature of the FOV. It also shows that within the overlapping coverage area, a region of high interference is created, where the SINR is low. The area of this region can be reduced by decreasing the number of deployed LiFi APs or the FOV of the light sources. However, this increases the coverage hole, and hence the handover-failure rate. Fig. 2(b) shows that the users at the edge of a room experience low SINR, as the received signal is low, and they are close to interfering femtocells.

The optical and RF spectrum do not interfere with each other; this is the most important advantage to be gained by deploying RF and optical wireless systems together. A hybrid LiFi/femtocell system can effectively solve the abovementioned limitations of individual LiFi and femtocell systems. Fig. 3(a) shows such a hybrid system. A femtocell deployed in a room overcomes the coverage-hole limitation of the LiFi network. As a result, there is no handover failure whenever a user moves inside a room. Moreover, overlapping LiFi users can be moved to the femtocell network, and thus, the SINR for these users is improved. The users at the edges of the room are covered by the LiFi network, and hence, receive less RF interference. The femtocell-to-femtocell interference can be reduced by employing a frequency-reuse technique among the femtocells. Moreover, by shifting some femtocell users to LiFi networks, a few FAPs can be operated in idle mode, reducing the interference. Fig. 3(b) shows an example of frequency allocation among the femtocells. The allocated frequency band for the femtocells is divided into four sub-bands, yielding a frequency-reuse factor (FRF) of four.

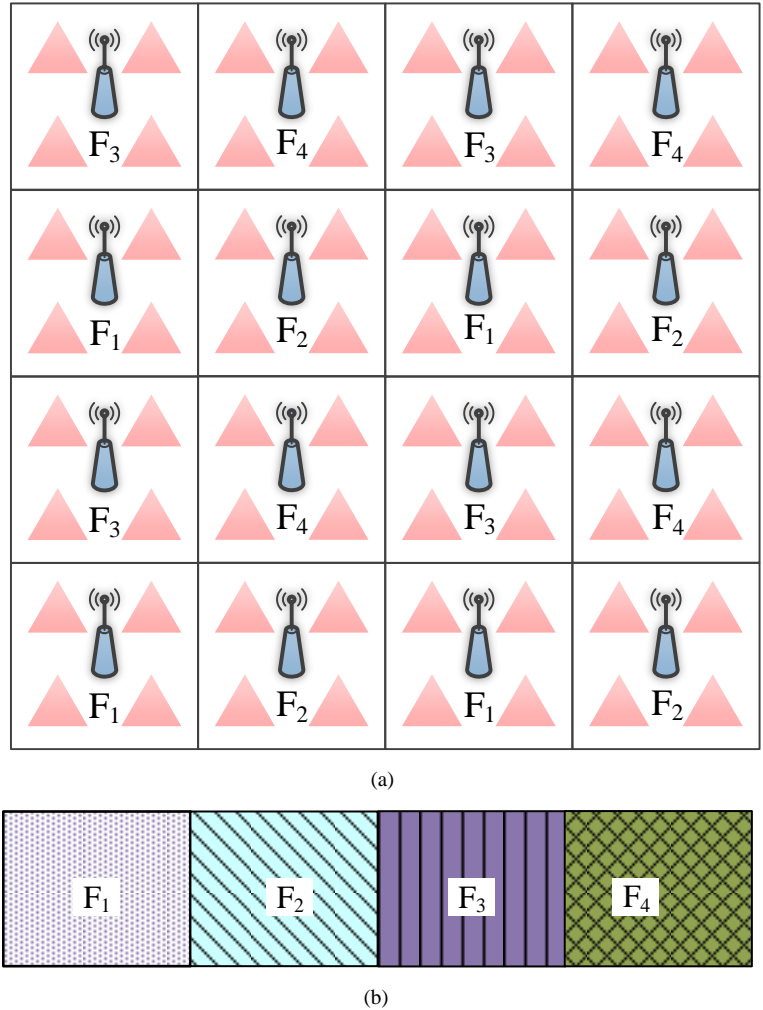

(a)

(b)

**Fig. 3.** Hybrid LiFi/femtocell system. (a) Hybrid scenario. (b) Frequency allocation among femtocells.

### 3.2 Analytic *Hierarchy Process (AHP) for Network Selection*

Network selection among the LiFi APs and femtocells can be accomplished using AHP [37], [38], which is a multi-criterion, decision-making (MCDM) method for developing a hierarchy among decision criteria and ranking the available alternatives. The main issue of an MCDM problem is to formulate a decision matrix with multiple alternatives and multiple decision criteria. AHP establishes a hierarchy-based, sub-problem approach. Fig. 4 shows the hierarchical structure of AHP, which relates the elements of one level to the level directly below it, thus creating a relation from the bottom level all the way to the topmost level. Hence, the global weights of the alternative LiFi and femtocell networks are found by multiplying the final two matrices:

$$\begin{bmatrix} a_{11} & a_{12} & a_{13} & a_{14} & \cdots & a_{1N} \\ a_{21} & a_{22} & a_{23} & a_{24} & \cdots & a_{2N} \end{bmatrix} \times \begin{bmatrix} W_1 \\ W_2 \\ \vdots \\ W_N \end{bmatrix} = \begin{bmatrix} R_1 \\ R_2 \end{bmatrix} \quad (8)$$

where $a_{1i}$ and $a_{2i}$ denote the values of the alternative LiFi and femtocell networks, respectively, with respect to the $i^{th}$ criterion; $W_i$ is the weight of the $i^{th}$ criterion; and $R_1$ and $R_2$ represent the weights of the LiFi and femtocell networks, respectively.

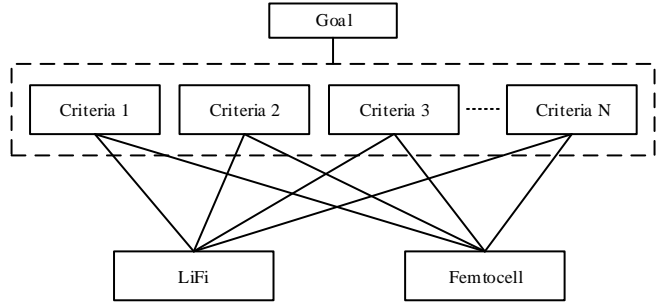

**Fig. 4** Hierarchical structure of AHP for network selection between LiFi and femtocell.

### 3.3 Zoning the Coverage

Due to the simultaneous presence of both LiFi and femtocell networks, the signal quality—in terms of received signal strength and interference levels—at different positions in a room is not the same. Therefore, for the proposed hybrid system, we introduce the zoning of coverages based on different criteria. Fig. 5 shows the zoning of wireless coverage inside a room. We propose four zones inside the room. Zone 1 consists of an area where only femtocell coverage is available. This zone plays a very important role in supporting LiFi-to-LiFi handover. Zone 2 consists of the central LiFi coverage areas. A circle is drawn to exclude the overlapping LiFi coverages. The LiFi edge coverages, discounting the overlapped LiFi coverages, are considered as Zone 3. Finally, the overlapping LiFi coverages are considered as Zone 4. This zoning of the wireless coverage can help to efficiently handle both users inside the room and the available resources, providing a better QoS level.

It is very important to know the zone-coverage information for analyzing different QoS parameters. The number of deployed LiFi APs within a room of size $a \times b$ is

$$N_{L,\min} = \left\lfloor \frac{a}{2r} \right\rfloor \times \left\lfloor \frac{b}{2r} \right\rfloor \tag{9}$$

$$N_{L,\max} = n_x n_y = \left\lfloor \frac{a}{2r} + 1 \right\rfloor \times \left\lfloor \frac{b}{2r} + 1 \right\rfloor \tag{10}$$

where $N_{L,\min}$ and $N_{L,\max}$ are, respectively, the minimum and maximum numbers of deployed LiFi APs within the room; $r$ is the coverage range of an LiFi AP; $a$ is the length of the room in the X-direction; $b$ is the length of the room in the Y-direction; $n_x$ is the number of deployed LiFi APs in the X-direction; and $n_y$ is the number of deployed LiFi APs in the Y-direction.

The distance between two LiFi APs in the X-direction is

$$d_x = \frac{a}{\left\lfloor \frac{a+2r}{2r} \right\rfloor} \tag{11}$$

The maximum overlapped distance between two LiFi APs in the X-direction is

$$l_x = \frac{2rn_x - a}{n_x} \tag{12}$$

The distance between two LiFi APs in the Y-direction is

$$d_x = \frac{b}{\left\lfloor \frac{b+2r}{2r} \right\rfloor} \tag{13}$$

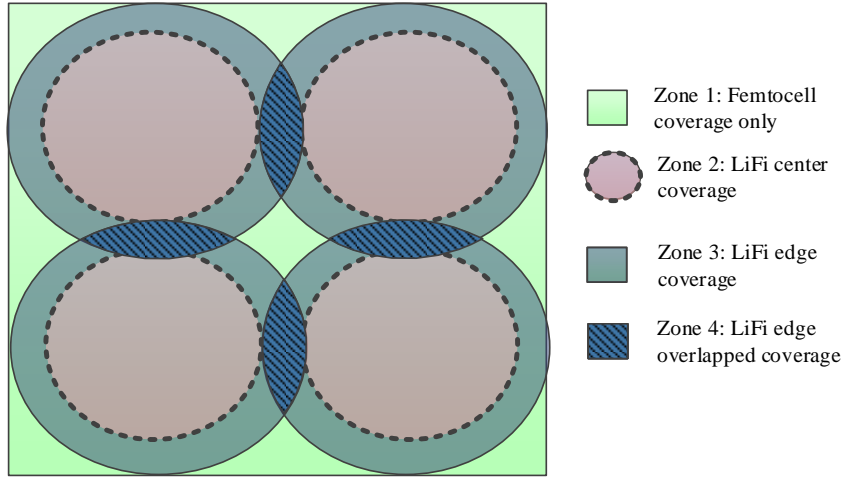

**Fig. 5.** Zoning of wireless coverage provided by hybrid LiFi/femtocell networks.

The maximum overlapped distance between two LiFi APs in the Y-direction is

$$l_y = \frac{2rn_y - b}{n_y} \quad (14)$$

The total area of Zone 1 is

$$A_{Z1} = ab - n_x n_y \pi r^2 + 4 n_x n_y \left[ \int_{r-\frac{l_x}{2}}^{r} \sqrt{r^2 - x^2}\, dx + \int_{r-\frac{l_y}{2}}^{r} \sqrt{r^2 - x^2}\, dx \right] \quad (15)$$

The total area of Zone 2 is

$$A_{Z2} = n_x n_y \pi \left( r - \frac{l_z}{2} \right)^2 \Big|_{l_z = \max(l_x,\, l_y)} \quad (16)$$

The total area of Zone 4 is

$$A_{Z4} = 4 n_x n_y \left[ \int_{r-\frac{l_x}{2}}^{r} \sqrt{r^2 - x^2}\, dx + \int_{r-\frac{l_y}{2}}^{r} \sqrt{r^2 - x^2}\, dx \right] \quad (17)$$

The total area of Zone 3 is

$$A_{Z3} = n_x n_y \pi r^2 - A_{Z2} - A_{Z4} \quad (18)$$

The probability that the number of users in the $i^{th}$ Zone is $m$ is

$$P_{m,zi} = \binom{p}{m} \left( \frac{A_{zi}}{ab} \right)^m \left( 1 - \frac{A_{zi}}{ab} \right)^{p-m} \quad (19)$$

where $p$ is the total number of users within the total hybrid LiFi/femtocell coverage area.

### 3.4 Call Admission and Interference Management

The call admission policy plays an important role in providing QoS support. We prefer to carry real-time (RT) voice calls on a femtocell network, as voice users require more mobility support. Fig. 6 shows the call admission policy for newly originating calls in a hybrid LiFi/femtocell system. Whenever a new call arrives, the system checks the zone first. If the call originates in Zone 1, then there is no choice of LiFi network and the call is accepted by the FAP. If the call is a RT voice call, then also it is accepted by the FAP. If the user is in Zone 4, the call is again accepted by the FAP. If the user is in Zone 3, the system initially checks whether the FAP is in idle mode. If the FAP is in idle mode, then the call is accepted by the LiFi network; otherwise, it is accepted by the FAP network. An FAP or LiFi accepts a call if resources in the respective network are available.

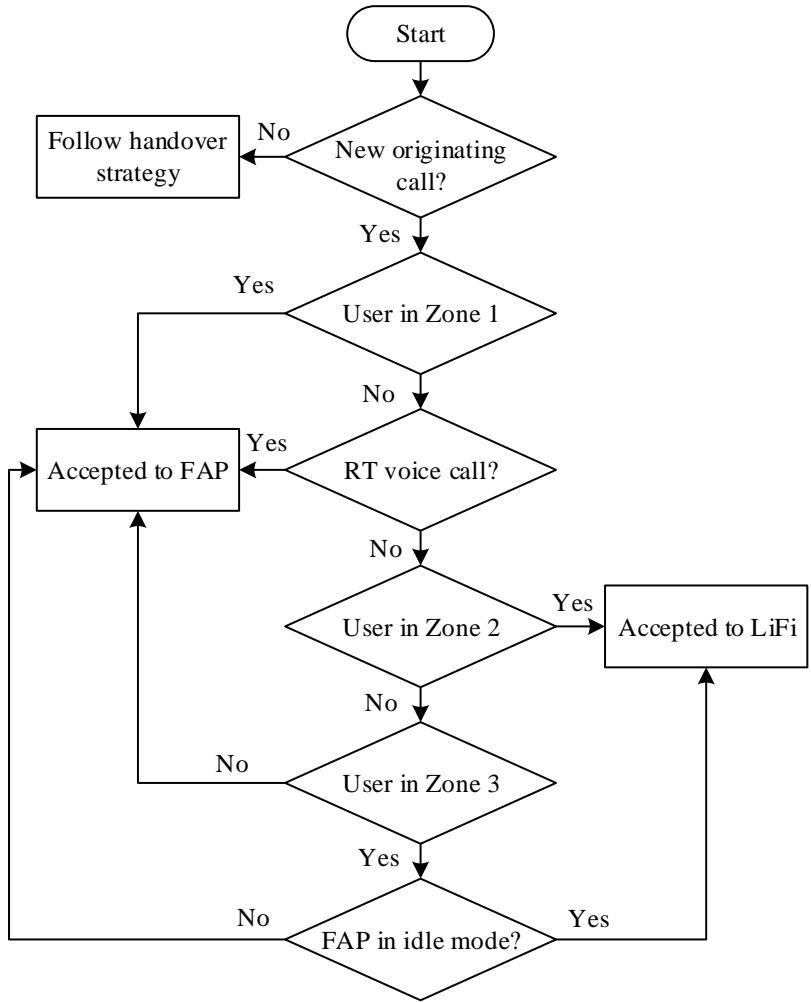

**Fig. 6.** Call admission policy for newly originating calls.

Seamless mobility is the most important issue for mobile users. Only a hybrid LiFi/femtocell system can ensure the seamless movement of users. Fig. 7 shows the admission policy for handover calls. The system initially checks whether the connected network is LiFi or FAP. For a LiFi-connected user, the possible handovers are LiFi-to-LiFi or LiFi-to-femtocell. If the user enters Zone 3 or Zone 1, then the user is handed over to the FAP. However, if the user enters Zone 4, the system considers signals received from both the serving LiFi and target LiFi, as well as the staying time in this zone. If the received signal $S_t$ from the target LiFi is greater than signal $S_s$ from the serving LiFi, the call is handed over to the target LiFi. If $S_s \geq S_t$ for a threshold time $T_h$ within Zone 4, the user remains with the serving LiFi. After passing the threshold time, the user is handed over to the FAP. However, if the user is connected with the FAP and receives a signal from LiFi, then there is a possibility of an FAP-to-LiFi handover. If the user moves to Zone 2, then the user is directly handed over to the LiFi network. If the user moves to Zone 3 and stays for more than the threshold time $T_{h1}$, the call is handed over to LiFi; otherwise, it stays with the FAP.

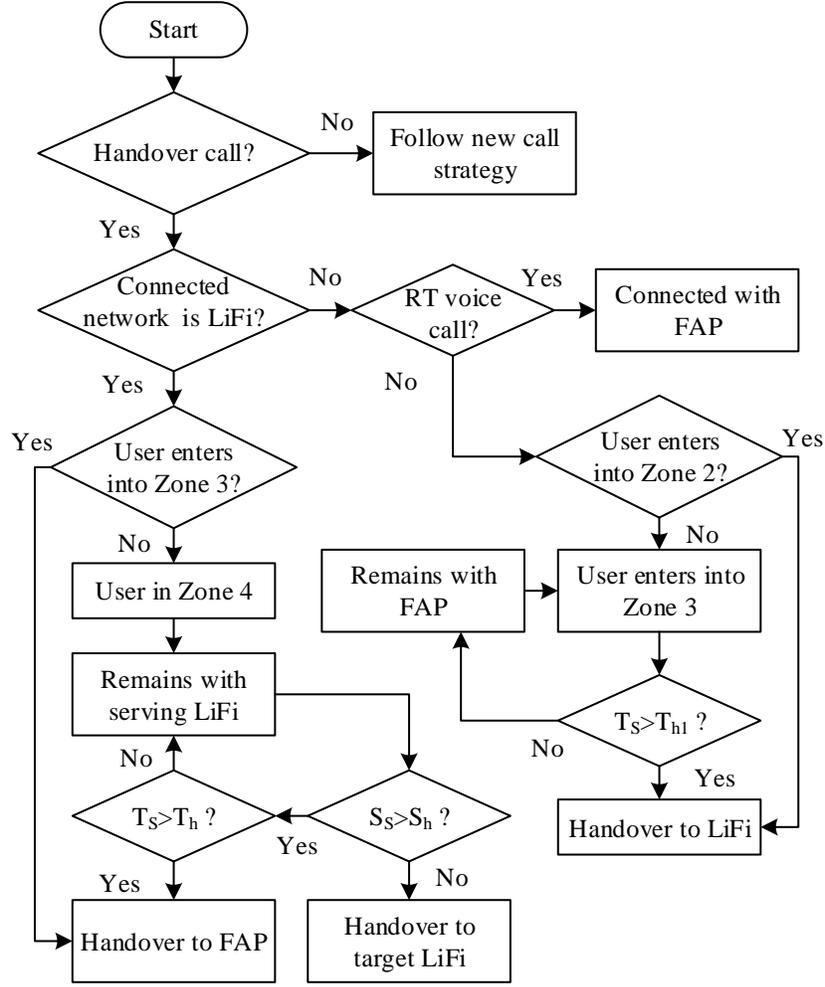

**Fig. 7.** Call admission policy for handover calls.

We propose the following operating-mode selection for the FAP. The idle mode of an FAP reduces the interference with neighboring FAPs. Fig. 8 shows the proposed FAP mode-selection procedure. If no user is connected with an FAP, then it goes into idle mode. Shifting an FAP user into LiFi also increases the number of idle FAPs in an area. If only one user is connected with an FAP, and if this user stays in Zone 3, then the user is moved to the LiFi network and the FAP changes its mode to idle.

The probability that an FAP can be switched to idle mode is

$$P_{F,I} = \sum_{k=0}^{1} \binom{p}{k} \left( \frac{A_{z1} + A_{z3}}{ab} \right)^{k} \left( \frac{A_{z2} + A_{z4}}{ab} \right)^{p-k} \qquad (20)$$

where $p$ is the total number of users within the total hybrid LiFi/femtocell coverage area.

### 3.5 Handover Call Flow

To date, no researcher has proposed an effective and complete handover call flow for LiFi network deployment. Here, we propose a complete handover call flow protocol for a hybrid LiFi/femtocell network. Three types of handovers can occur in this system: (i) LiFi-to-femtocell, (ii) femtocell-to-LiFi, and (iii) LiFi-to-LiFi.

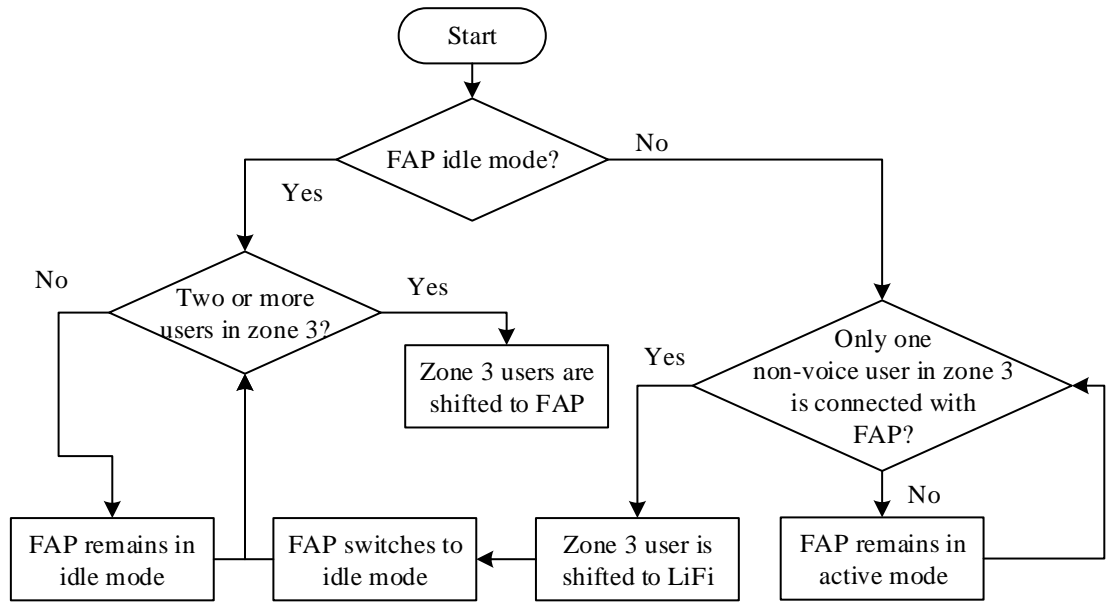

**Fig. 8.** FAP idle-mode selection.

The user equipment (UE) needs to select an appropriate target AP from among the FAP and LiFi APs. In addition, the interference level must be considered in making the handover decision. Handover from LiFi-to-femtocell is simpler compared to others, as only a single option for handover is available. The basic steps for handover include signal-level measurement, appropriate AP selection for the handover, handover decision, and handover execution.

### 3.5.1 LiFi-to-Femtocell Handover

Fig. 9 shows the detailed call flow procedures for a LiFi-to-femtocell handover in a hybrid LiFi/femtocell network. If a UE senses that the optical signal from a currently connected LiFi AP is decreasing, the UE sends a report to its connected LiFi AP (steps 1 and 2). The UE searches for signals from the neighboring LiFi APs and the overlaid FAP (step 3). The UE and the serving LiFi AP together decide on the best AP for the handover (step 4). The UE checks pre-authentication with the selected FAP (step 5). On the basis of pre-authentication and the received signal levels, the UE and serving LiFi AP together decide on the handover to the FAP (step 6). The serving LiFi AP starts the handover procedure by sending a handover request to the FAP through the gateway (GW) (steps 7 and 8). Call admission control (CAC) is performed to check whether the call can be accepted (step 9). Then, the FAP responds to the handover request (steps 10 and 11). Steps 12–14 are used to set up a new link between the GW and FAP, and the packet data are forwarded to the FAP (step 15). The UE re-establishes a radio channel with the FAP, detaches from the serving LiFi, and synchronizes with the FAP (steps 16–20). The UE sends a "handover complete" message to the GW to inform it that the UE has already completed handover and has synchronized with the target FAP (steps 21 and 22). Then, the serving LiFi deletes the old optical link with the GW (steps 23−25). The packets are then sent to the UE through the FAP.

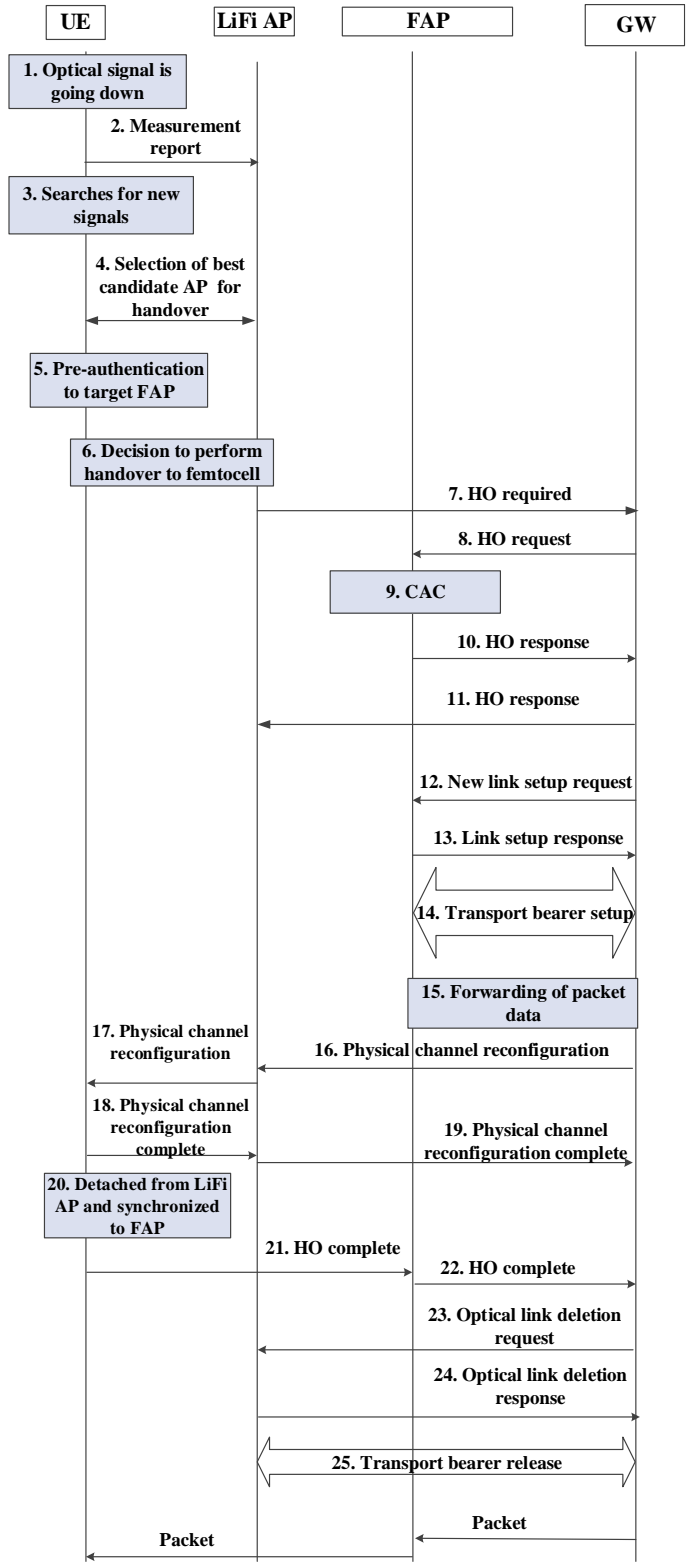

**Fig. 9.** Call flow for a LiFi-to-femtocell handover.

### 3.5.2 Femtocell-to-LiFi Handover

This is an optional handover, as both femtocell and LiFi networks are available to serve the user. In this handover, the UE needs to select an appropriate target LiFi AP. In addition, the interference level must be considered in making the handover decision. Fig. 10 shows the call flow procedures for a femtocell-to-LiFi handover. Whenever the UE detects an optical signal from an LiFi AP, it sends a measurement report to the connected FAP (steps 1 and 2). The UE and FAP together select the best possible candidate LiFi AP for the handover (step 3). The UE checks pre-authentication with the selected LiFi AP (step 4). On the basis of the pre-authentication and the received signal levels, the UE and FAP together decide on the handover to the LiFi AP (step 5). The FAP starts the handover procedure by sending a handover request through the GW (steps 6 and 7). The target LiFi AP checks the user authorization (steps 8 and 9). The target LiFi AP performs CAC and interference-level comparisons to admit a call (step 10), and then responds to the handover request from the FAP (steps 11 and 12). A new link is established between the GW and target LiFi AP (steps 13–15), and the packet data are forwarded to the target LiFi AP (step 16). Now, the UE re-establishes an optical channel with the target LiFi AP, detaches from the serving FAP, and synchronizes with the target LiFi AP (steps 17–21). The UE sends a "handover complete" message to the GW to inform it that the UE has already completed the handover and has synchronized with the target LiFi AP (steps 22 and 23). Then, the FAP deletes the old link with the GW (steps 24–26), and the packets are forwarded to the UE through the target LiFi AP.

### 3.5.3 LiFi-to-LiFi Handover

This handover is very important for ensuring the QoS level. Fig. 11 shows the detailed call flow procedures for a LiFi-to-LiFi handover. If a UE detects the optical signal strength decreasing, it sends a report to the connected target LiFi AP (steps 1 and 2). The UE searches for signals from neighboring APs (step 3). The UE and serving LiFi AP collectively select the best AP for the handover (step 4). The UE performs pre-authentication with the selected LiFi AP (step 5). On the basis of the pre-authentication and received signal levels, the UE and serving LiFi AP together decide on handover to the target LiFi AP (step 6). The serving LiFi AP starts the handover procedure by sending a handover request to the target LiFi AP through the GW (steps 7 and 8). The target LiFi AP checks the user authorization (steps 9 and 10). The target LiFi AP performs CAC and interference-level comparisons for the handover call (steps 11). Then, the target LiFi AP replies to the handover request (steps 12 and 13). A new link is established between the GW and target LiFi AP (steps 14–16), and the packet data are forwarded to the target LiFi AP (step 17). The UE then re-establishes a channel with the target LiFi AP, detaches from the serving LiFi AP, and synchronizes with the target LiFi AP (steps 18–22). The UE sends a "handover complete" message to the GW to inform it that the UE has already completed handover and has synchronized with the target LiFi AP (steps 23 and 24). Then, the serving LiFi AP deletes the link with the GW (steps 25–27), and the packets are forwarded to the UE through the target LiFi AP.

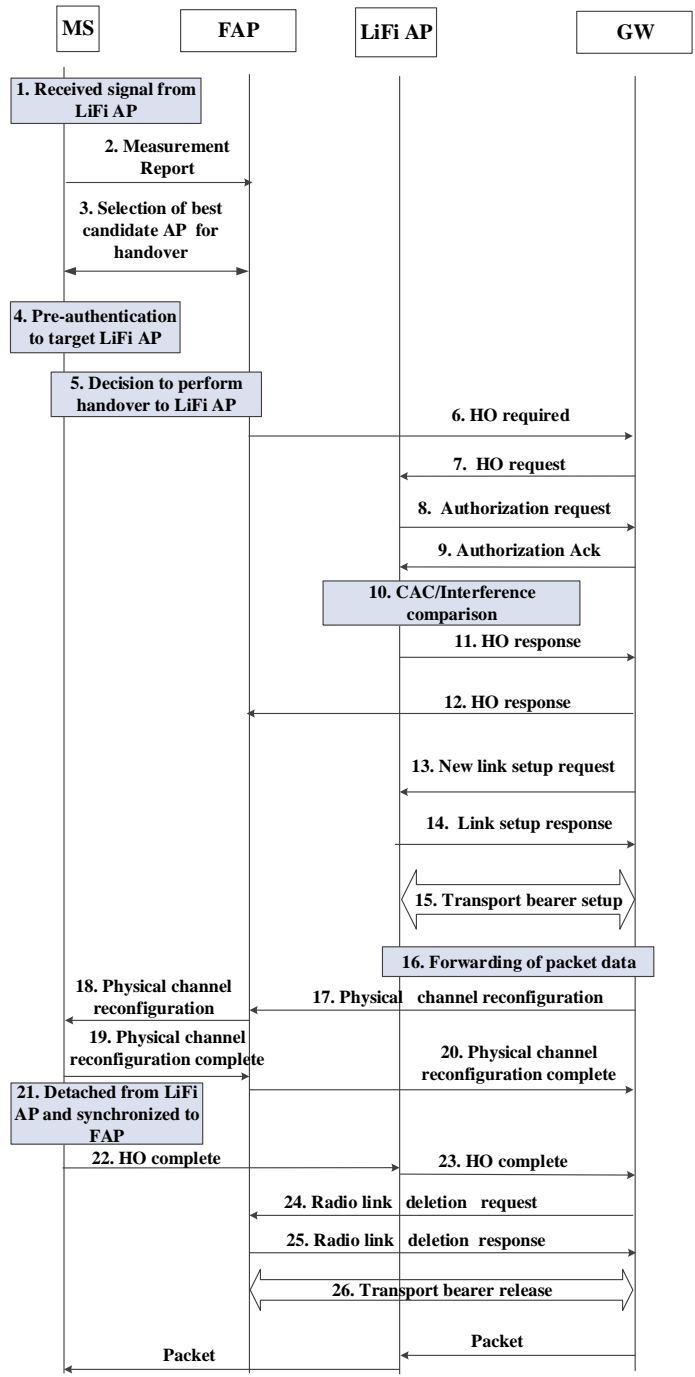

**Fig. 10.** Call flow for a femtocell-to-LiFi handover.

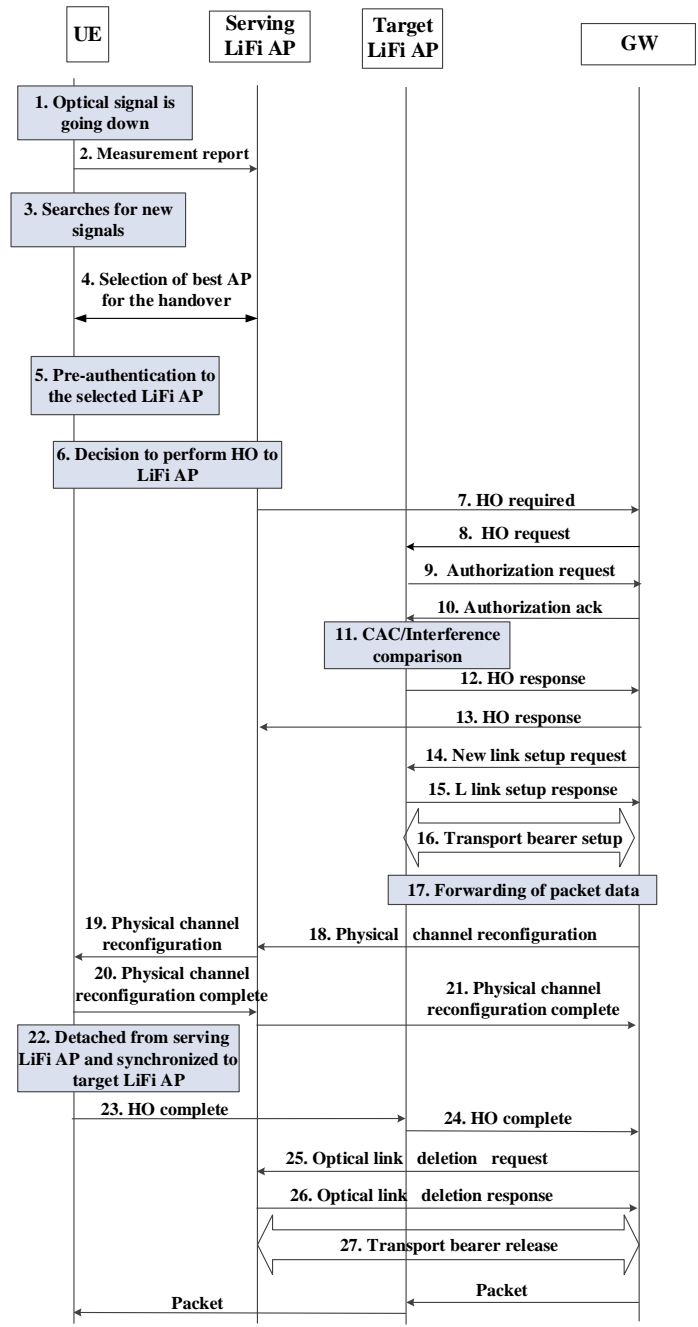

**Fig. 11.** Call flow for a LiFi-to-LiFi handover.

## 4. Integrated RF/Optical Wireless in Transportation

Integrated RF/optical networks can play an important role in improving the QoS for wireless connectivity in transportation. In this section, we discuss several issues regarding wireless connectivity when using an integrated architecture for vehicles.

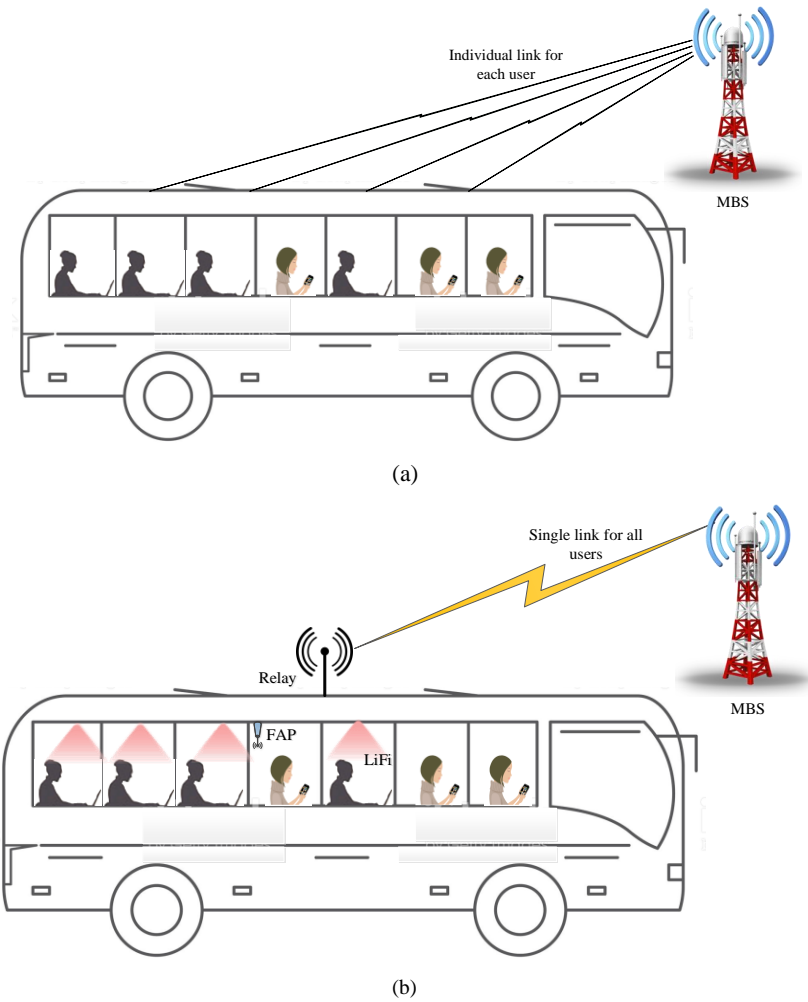

**Fig. 12.** A scenario for wireless connectivity in public transportation. (a) Connectivity using a macrocellular network only. (b) Connectivity using an integrated network.

### 4.1 Improved Signal Quality

The users inside a vehicle are usually connected to a macrocell network through direct individual connectivity. Hence, the transmitted signal from MBS must penetrate the vehicle body, which causes a significant reduction in the signal level. Here we propose an integrated RF/optical system that improves the service quality for users inside a vehicle. A relay is installed outside the vehicle to receive the macrocell signal. Wireless connectivity for the inside LOS users are provided using LiFi APs and for NLOS as well as voice users are provided using FAPs. The data users use LiFi, while the voice users use femtocells for their connectivity. The femtocell network is better for voice service, as it can cover a larger area than LiFi, and the FAP can provide NLOS communication. Hence, the users enjoy better signal quality, as they are closer to their APs. Fig. 12 illustrates wireless connectivity in a bus. Fig. 12(a) shows the connectivity using a macrocell network only, with each connection based on a distinct link. Fig. 12(b) shows the connectivity using an integrated network. MBS has only one link for all users inside the vehicle. Inside the bus, the simultaneous presence of LiFi and femtocell forms a hybrid system, whereas for a LiFi user, the link to MBS forms a coexisting RF/optical wireless system.

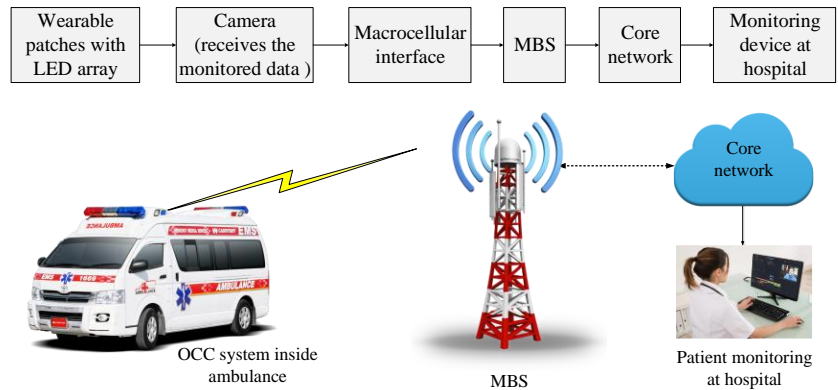

**Fig. 13.** Scenario illustrating patient monitoring in an ambulance from a remote location.

### 4.2 Ambulance-Patient Monitoring

Sometimes, it takes several hours to move a patient from one city to another by ambulance. It is very important to monitor a serious patient during this movement. RF technologies for eHealth solutions have several disadvantages. Especially for healthcare solutions, RF can cause serious electromagnetic interference. Moreover, RF is harmful to patients. Hence, instead of RF sensors, an OCC system can be used for simultaneously collecting multiple sensors data. In addition, a video camera in the OCC system can monitor the patient. Therefore, an OCC system can be installed inside the ambulance to collect the monitored data from wearable sensors. These sensors [39]-[41] have LEDs or LED arrays that act as the physical transmitters for the OCC system. The collected data are forwarded to the hospital or the concerned doctor using macrocellular networks. Fig. 13 shows a scenario illustrating patient monitoring in an ambulance from a remote location using a coexisting RF/optical wireless system. The patient's condition can be continuously monitored by the remote doctor, and he/she can guide the treatment accordingly from a remote location.

### 4.3 Tracking a Vehicle

Currently, global positioning system (GPS) is widely used for tracking vehicles. However, the GPS system is costly, and cannot provide services in tunnels or underground. Thus, our proposed integrated optical/RF wireless system can be a good complementary option to the GPS system. Fig. 14 shows such an example of a vehicle-tracking system. Every vehicle has a unique identification (ID). An OCC system is installed with each overhead traffic signal. The camera receives the ID from the vehicle whenever it is within the communication range. This ID, together with the location information for the infrastructure and the distance of the vehicle from the infrastructure, is sent to the server, which calculates the precise location of the vehicle. The passenger can easily access this information through a smartphone or other smart device and can locate the vehicle. At a bus stop, LiFi or WiFi connectivity can provide location information. This system can also be applied to localize a train without using GPS.

### 4.4 Group-Handover Management

The use of coexisting wireless technologies can provide a good choice for group-handover provisioning. Whenever individual wireless links are provided for each user inside a vehicle, individual handovers must occur if a vehicle moves from one macrocellular network coverage area to another. However, the support of inside users through LiFi and/or femtocell networks can avoid such individual handovers. Only a single group handover needs to be performed, saving both the amount of signaling and the bandwidth. In a group handover, only one link exists between MBS and the relay situated outside the vehicle. The outside relay and inside LiFi/FAP are connected through a wired line.

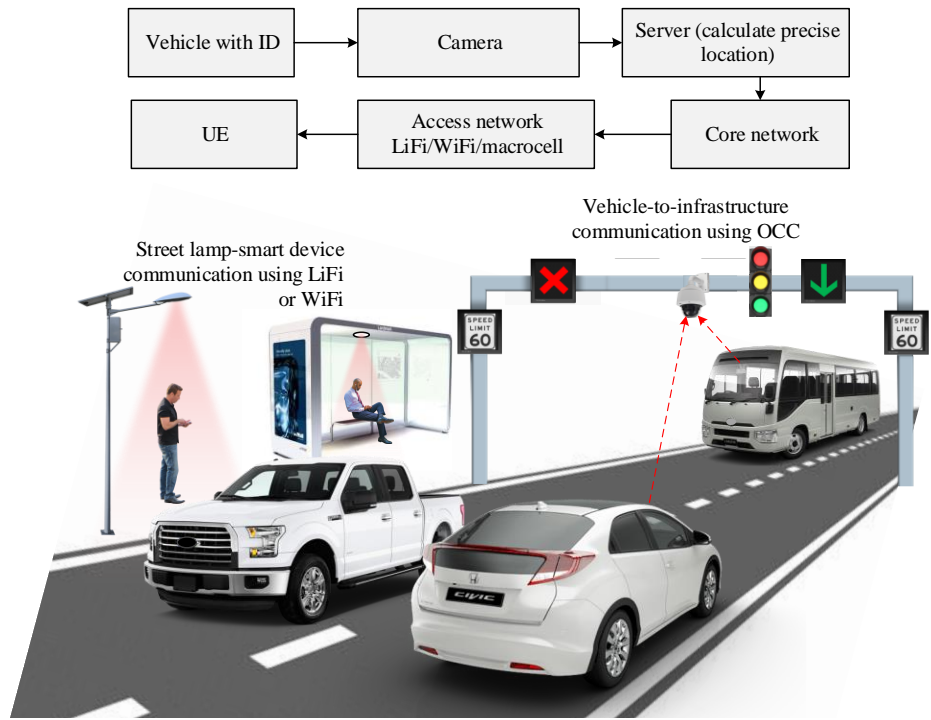

**Fig. 14.** Scenario illustrating a vehicle-tracking system based on an integrated optical/RF network.

**4.5 Improvement in Car-to-Car Link Reliability**

Hybrid RF and optical wireless technologies can also be used to improve the link reliability in car-to-car communications. A link reliability is the probability that the communication link between the transmitter and receiver is maintained without interruption. Various environments—e.g., excess interference caused by RF signals, foggy weather, longer distances between two vehicles, and NLOS communication—can affect car-to-car communications. Both RF and optical wireless technologies have performance imitations in such communication environments. The most important disadvantage for RF communication is the large amount of interference due to by nearby APs. On the other hand, OWC technologies are very sensitive to NLOS communication. A hybrid system can improve link reliability for car-to-car communications. Fig. 15 shows a few examples of improvement in link reliability for car-to-car communications that can be obtained using hybrid RF and optical wireless technologies. The first example concerns NLOS communication due to non-visual, bad-weather conditions. Accident information from the front car cannot be conveyed to the back car using an OCC system because the receiver cannot detect the light transmitted from the front car. In this case, the alternative RF system is used to communicate between the cars. The second case shows an example where the distance between two cars is quite far out of the RF communication range. Hence, the cars cannot communicate with each other. An FSOC system can provide a long-distance link between the cars, which can solve this issue. FSO systems use laser technology for signal transmission and are widely operated in the near-infrared spectrum. Because of coherent nature in laser lights, they are concentrated and therefore, directed forward for long distances. The third case shows a situation in which the front car turns left or right, creating an NLOS situation. At that time, the OWC communication link is interrupted. In this scenario, RF-based communication can provide better connectivity. Conversely, congestion due to the presence of many vehicles on the road can produce too many RF signals, causing serious interference, and link reliability is degraded. In this case, we can replace a few RF links with optical wireless links. The alternative optical wireless links improve the quality of communication. Another option is to use a smart antenna that can automatically convert from omnidirectional to directional. In this case, the front-car-to-back-car link can be performed using an RF communication network, while the back-car-to-front-car link can be maintained using an OCC system, or vice versa.

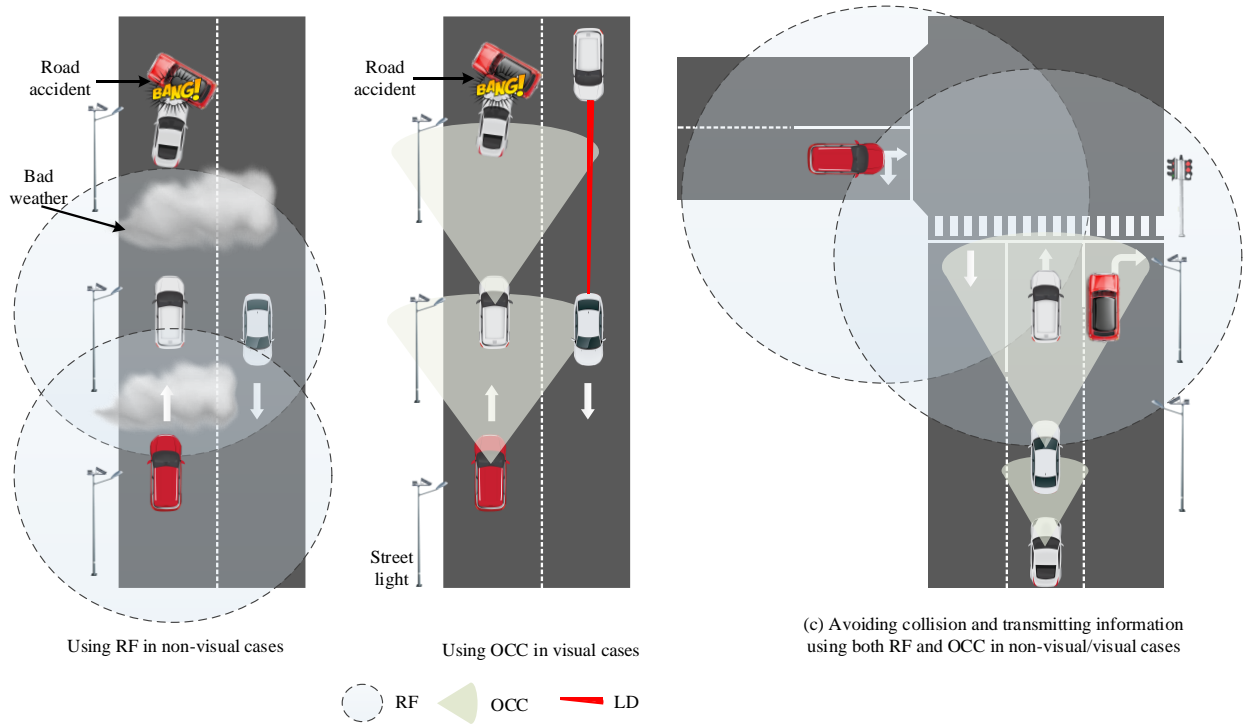

**Fig. 15.** Scenarios illustrating improved link reliability in car-to-car communications using hybrid RF/optical wireless technologies.

## 5. Performance Evaluation

In this section, we evaluate the performance of a integrated RF/optical wireless system. The results show the features and improvements provided by the proposed scheme, which ensures a better QoS level for the user. Table I summarizes the fundamental parameters that we used to define the performance analysis.

### 5.1 Indoor Hybrid LiFi/Femtocell Solution

Both LiFi and femtocell users benefit from the coexistence of LiFi and RF femtocell networks. In particular, seamless handovers can be supported for the LiFi users with the help of femtocells. On the other hand, users can enjoy high-data-rate services through LiFi connectivity. Also, femtocell users may experience high interference due to the large-scale deployment of femtocells; in this case, users receiving low SINR can be shifted to LiFi networks. To carry out the performance analysis of the proposed hybrid network, we consider a room of size 24 m × 24 m, with nine LiFi APs. We consider only femtocells within a 100 m range of the reference FAP.

We propose to employ mode switching for the FAPs, with an FAP switched to idle mode if it is not needed to serve any user. Because idle-mode transmission power is very low, interference is reduced. Fig. 16 shows the probability that an FAP is in active mode, as a function of the change in the number of active users in the room. There is a high probability that the FAP is in idle mode when the number of active users in the room is low. This is because a large portion of the room is covered by good-quality LiFi coverage, and the users within this area are served by the LiFi system. Hence, the idle-mode technique improves the performance of the femtocell system.

Table I. Parameter values used in the performance analysis

| RF System | |
|---|---|
| Center frequency for macrocell and femtocell | 1800 MHz |
| Transmitted signal power by the MBS for edge zone | 46 dBm |
| Transmitted signal power by the FAP | 7 dBm |
| Noise power density | -174 dBm/Hz |
| Distance between two macrocells | 1000 m |
| Building wall penetration loss | 20 dB |
| Vehicle wall penetration loss | 10 dB |
| Height of the MBS | 50 m |
| Height of the FAP | 3 m |
| Height of the indoor/vehicle UE | 1 m |
| Distance power loss coefficient for femtocells | 28 |
| **Optical system** | |
| Height of the LiFi AP | 3 m |
| LiFi coverage | 5 m |
| Transmitted optical power per LiFi AP | 6 W |
| Optical modulated bandwidth | 20 MHz |
| Area of PD | 1 $cm^2$ |
| Optical filter gain | 1 |
| Optical to electrical power conversion efficiency | 0.53 A/W |
| Half intensity radiation angle | 60 degree |
| Receiver FOV semi-angle | 90 degree |
| Refractive index | 1.5 |
| Noise power spectral density | $10^{-21}$ $A^2$/Hz |

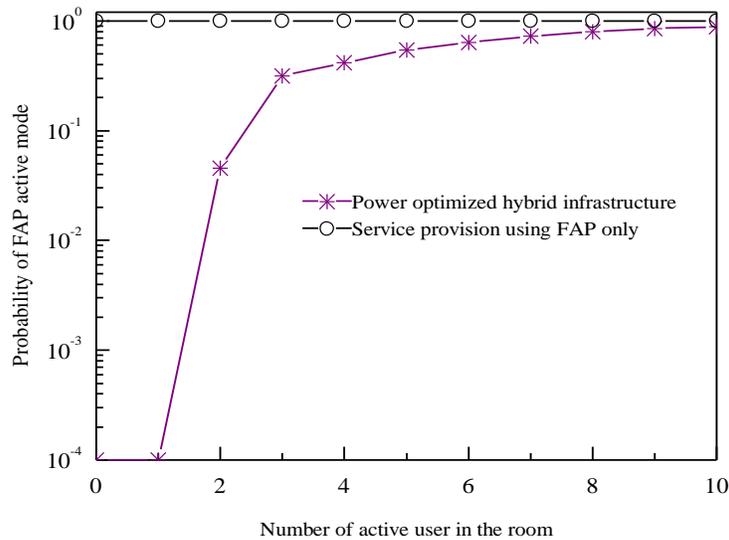

**Fig. 16.** Probability that an FAP is in idle mode, as a function of change in the number of active users.

Our proposed hybrid network also improves the SINR performance for femtocell users. Fig. 17 shows an analysis of SINR levels for a femtocell user. We take the user distance from the reference FAP to be 8 m, and assume that 50 FAPs are deployed within a 100 m range. Our proposed scheme outperforms a pure femtocell network system because we properly allocate power to FAPs and employ idle-mode operation. We consider two cases of frequency allocation among the femtocells: (i) FRF 1 and (ii) FRF 4. For both cases, the performance is significantly enhanced by the proposed hybrid network. The transmitted powers of the idle-mode FAPs do not interfere with those of neighboring FAPs, and we thus achieve a better SINR level. Hence, we can also achieve better capacity [bps/Hz] and a lower outage rate for the femtocell system.

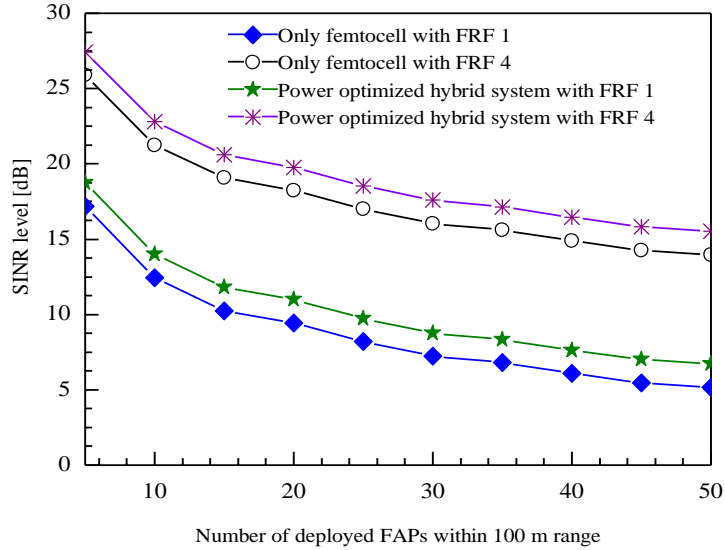

**Fig. 17.** SINR performance evaluation for femtocell users.

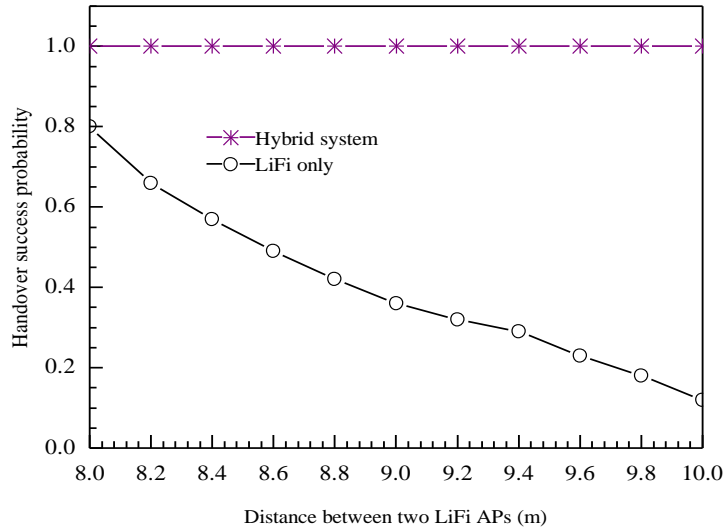

**Fig. 18.** Comparison of handover success performance.

Our proposed hybrid system also smooths the handover process in LiFi network deployment because the LiFi system provides mobility support. However, the LiFi coverage is comparatively small, and a coverage hole exists in the LiFi system. Fig. 18 shows improvement in handover performance for the proposed hybrid LiFi/femtocell system. Whenever the distance between two LiFi APs is greater, there is a higher probability of the user moving into an area where a LiFi coverage hole exists. Therefore, the handover from LiFi-to-LiFi is dropped. The performance improves when the distance between two LiFi APs is decreased. However, this causes an increase in the area of Zone 4, and thus, increases interference. The provision of femtocell coverage in hybrid system provides the facility to overcome the LiFi coverage hole and thus, maintains 100% handover success probability.

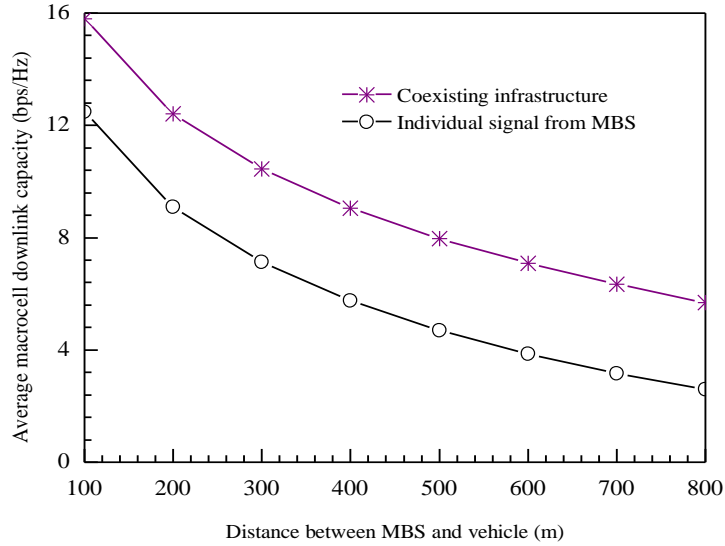

**Fig. 19.** Downlink capacity comparison of macrocellular networks for vehicle users.

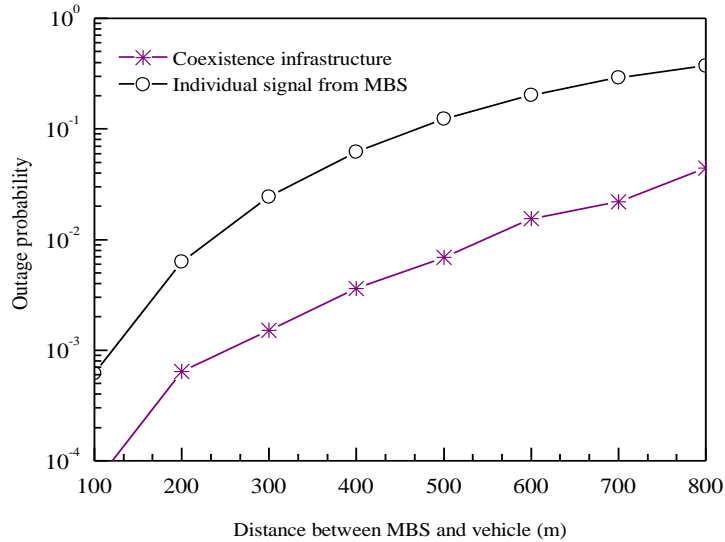

**Fig. 20.** Outage probability comparison for vehicle users.

### 5.2 Integrated Optical/RF Wireless System for Transportation

An integrated RF/optical system can also significantly improve wireless connectivity performance for transportation. We assume the threshold value of SNIR sensitivity for a users' receiver and an outside relay receiver to be 9 and 5 dB, respectively. Fig. 19 shows that the integrated infrastructure improves the downlink capacity significantly for vehicle users. This is because the users are close to the APs and the proposed architecture avoids vehicle-wall-penetration loss. Hence, the users receive better signal quality and the overall capacity is improved. Fig. 20 shows that our proposed hybrid model significantly reduces the connection-outage probability for vehicle users. This is because the users avoid the vehicle-penetration loss and that the outside relay receiver is less sensitive than the users' receivers. The hybrid RF/optical wireless infrastructure also improves the reliability of car-to-car communications. Fig. 21 shows an analysis of the link reliabilities. We assume a U-turn loop of radius 10 m, with the average speed of a vehicle crossing the loop being 40 km/h. The coverage range of the RF signal is taken to be 30 m. The figure clearly shows that the hybrid infrastructure improves the link reliability.

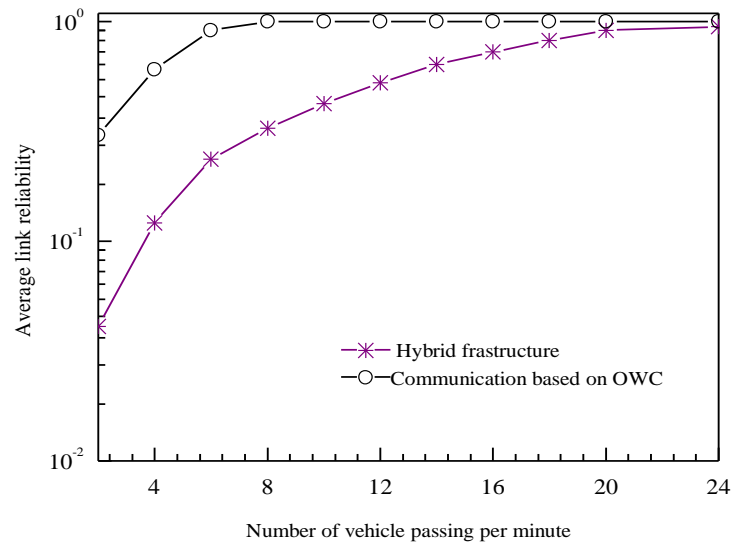

**Fig. 21.** Comparison of average link reliability for transportation.

## 6. Conclusions

Next-generation communications will require heterogeneous RF and optical wireless networks to meet high-data-rate demands. OWC is a promising solution for future wireless connectivity, as RF-based wireless technologies are not sufficient to fulfill the increasing demands. Both RF and optical wireless technologies have advantages and limitations, but their integration provides the best solution for overcoming these limitations. The RF and optical spectrum do not interfere with each other, which is the most important advantage of coexisting infrastructure. In this paper, we have discussed integrated RF/optical wireless systems for indoor and transportation usage, considering improvements in mobility support, call admission policy, interference management, capacity, and link reliability. We have described in detail hybrid LiFi/femtocell networks for indoor use. We have introduced zoning of the indoor coverage for the call admission policy and have considered the details of handover call flow, interference management, interface selection, and handover support. For transportation, we have discussed scenarios for services to users inside a vehicle, remote monitoring of ambulance patients, vehicle tracking, and car-to-car communications. We have considered macrocell, femtocell, and WiFi as the RF systems and OCC, VLC, LiFi, and FSOC as the optical wireless systems for integrated RF/optical wireless applications in transportation. The performance results also show the necessity of such integrated RF/optical wireless systems. We anticipate that our proposed integrated RF/optical wireless approach will be very effective for future 5G heterogeneous wireless networks and beyond.